\documentstyle[12pt]{article}

\newcommand{\be}{\begin{equation}}
\newcommand{\ee}{\end{equation}}
\begin{document}
\begin{titlepage}
\begin{flushright}
{\bf HIP-2000-08/TH}\\
{\bf\it hep-th/0003270}\\
May 20, 2000
\end{flushright}
\begin{center}
{\Large \bf Field Theory on Noncommutative Space-Time\\  
and the Deformed Virasoro Algebra}\\

\vskip 1cm

{\large {\bf M. Chaichian} $^{a}$},\ \  
{\large{\bf A. Demichev} $^{a,b}$}\ \ and \ \ 
{\large{\bf P. Pre\v{s}najder} $^{a,c}$}

\vskip 0.2cm

${}^a$ High Energy Physics Division, Department of Physics,\\
University of Helsinki\\
{\small and}\\
Helsinki Institute of Physics,\\
P.O. Box 9, FIN-00014 Helsinki, Finland\\
${}^b$ Nuclear Physics Institute, Moscow State University,\\
119899, Moscow, Russia\\
${}^c$ Department of Theoretical Physics, Comenius University,\\
Mlynsk\'{a} dolina, SK-84215 Bratislava, Slovakia

\end{center}

\begin{abstract}

We consider a field theoretical model on the noncommutative cylinder which
leads to a discrete-time evolution. Its Euclidean version is shown to be
equivalent to a model on the complex $q$-plane. We reveal a direct link between
the model on a noncommutative cylinder and the deformed Virasoro algebra
constructed earlier on an abstract mathematical background.  As it was shown,
the deformed Virasoro generators necessarily carry a second  index (in addition
to the usual one), whose meaning, however, remained unknown.  The present 
field theoretical approach allows one to ascribe a clear meaning to  this
second index: its origin is related to the noncommutativity of the underlying 
space-time. The problems with the supersymmetric extension of the model on a
noncommutative super-space are briefly discussed.

\end{abstract}

{\it PACS:}\ \ 03.70\hfill{\it Keywords:}\hspace{3mm}
{\small Noncommutative space-time; Virasoro algebra}
\end{titlepage}

\section{Introduction}
The string theory and the noncommutative geometry are fundamental theories
possessing common goals: the elimination of the problems appearing in the
standard field theory, like ultra-violet divergences, and, perspectively,
a quantization of the gravity. In the former theory the elementary objects are
strings, i.e. they are not point-like, whereas from the latter the notion
of a point, as the elementary geometrical or physical entity, is eliminated
from the very beginning (for detailed discussion see books \cite{GSW},
\cite{Pol} and \cite{Con}). However, it appears that there are deep relations
between both approaches: the Yang-Mills theory on a noncommutative tori
\cite{CR} can be re-interpreted as a limiting model of the $M$-theory
\cite{CDS}.

In \cite{SW} the idea is discussed that the noncommutative space-time can
appear naturally in the particular low energy limit of the string theory,
as a direct consequence of a (constant) non-vanishing $B$-field. This fact
is closely related to deformation quantization \cite{Scho}. Conversely, the
deformation quantization \cite{Kon} can be interpreted in terms of a
topological string theory \cite{CF}. Here the noncommutative geometry appears
in the space-time (the target space of string coordinates).

In this article we shall analyze the different link between the string
theory and the noncommutative geometry relating the known deformation of
the Virasoro algebra with the noncommutative geometry on a string
world-sheet.

The Virasoro algebra is an infinite Lie algebra with generators $L_m$, $m
\in {\bf Z}$, satisfying commutation relations
\be
[L_m ,L_{m'} ]\ =\ (m-m')L_{m+m'} \ +\ \frac{c}{12} (m^3 -m)
\delta_{m+m',0} \ ,
\ee
where $c$ is a central element commuting with all $L_m$. In an unitary
irreducible representation $L^+_m = L_{-m}$ and $c$ is a real constant.
The Virasoro algebra is usually realized in terms of an infinite set of
(bosonic or fermionic) oscillators, and is closely related to the symmetry
properties of the 2D (conformal) field theory in question. The simplest such
representation, with $c=1$, is given by the Sugawara construction of $L_m$
in terms of an infinite set of bosonic oscillators $a_n$, $a^+_n = a_{-n}$,
$n=1,2,\dots$, satisfying the commutation relations
\be
[a_n ,a_m ]\ =\ n \delta_{n+m,0} \ ,\quad n,m \neq 0\ .
\ee
The Sugawara formula for $L_m$ reads
\be
L_m \ =\ \frac{1}{2} \sum_{k,n\neq 0} :a_k a_n : \delta_{k+n,m} \ ,\
\ m\in {\bf Z}\ .
\ee

The deformations of the Virasoro algebra are related to deformations of the
Sugawara construction (3). The oscillator realizations (with a nontrivial
central element) of these deformations of Virasoro algebra were constructed in
\cite{CP1}. They have been intensively studied in \cite{CP1}-\cite{CP2}, mainly
in connection with a formal deformations in the conformal and/or string field
theories. Later a developed mathematical structure was found, the
Zamolodchikov-Faddeev algebras, which proved to be a natural framework  for the
deformed Virasoro algebras \cite{LP}. Our construction in \cite{CP1} represents
a  paricular realization of a bosonization of a ZF algebra.  For a recent
review see \cite{Od}.

Our strategy is as follows. First we describe a model on a standard
(commutative) cylinder, in which framework the Virasoro algebra appears
naturally. Then we generalize the model to the noncommutative analog of the
cylinder. This leads to the discrete time evolution.  In Sec. 2 we summarize
briefly free scalar field theory on a commutative and noncommutative cylinder.
In Sec. 3 we describe the Euclidean version of the model and analyze its
symmetry properties.  We show that there is a particular model on a
noncommutative cylinder possessing, as a symmetry algebra, the deformed
Virasoro algebra proposed earlier. Thus, the Virasoro algebra on a complex
plane ${\bf C}$ is replaced by the deformed Virasoro algebra on a $q$-deformed
complex plane ${\bf C}_q$: there is a direct link between the noncommutative
geometry on a world-sheet and the deformation of the Virasoro algebra. This
link allows us to ascribe a clear physical meaning to the second index of the
deformed Virasoro generators  which necessarily appears in all abstract
generalizations of the Virasoro algebra. It turns out that appearance of the
additional index is a  straightforward consequence of the noncommutativity of
coordinates of the space-time ($q$-complex plane or, equivalently,
noncommutative cylinder) on  which the underlying field theory is defined. The
field theoretical origin  of the deformed Virosoro algebra can serve for the
better (physical) motivation and  understanding of its role as well as
understanding of all related constructions ($q$-strings, $q$-vertex operators
and  Zamolodchikov-Faddeev algebras). Now all physical intuition (based on a
notions like field action, quantization, Fock space, etc) can be used for
further extensions, new constructions, etc originated and related to the
q-deformed Virasoro algebras. In the last Sec. 4 we formulate the problem of a
supersymmetric extension of the model and add concluding remarks.

\section{Scalar field on a cylinder}
\subsection{Commutative case}
In this section, we first discuss a free scalar field on a standard
(commutative) cylinder $C$ (see, e.g., \cite{GSW}, \cite{Pol} and refs
therein), and then we generalize the model to the
noncommutative case. More detailed description of field-theoretical models
on a noncommutative cylinder can be found in \cite{CDP1}-\cite{CDP2}.

The cylinder $C$ which we identify with the set of points $C={\bf
R}\times S^1 =\{ x=(\rho\cos\phi ,\rho\sin\phi ,\tau )\in{\bf R}^3,\
\rho=\mbox{const} \}$.
If the function $f(x)$ on $C$ can be expanded as
\be
f(\tau ,\varphi ) \ =\ \sum_{k\in {\bf Z}} c_k (\tau ) e^{ik\varphi} \ ,
\ee
we introduce the standard integral on $C$ by putting
\be
I_0 [f]\ =\ \frac{1}{2\pi}\int_C d\tau d\varphi f(\tau ,\varphi)\ =\
\int_{\bf R} d\tau \ c_0 (\tau ) \ .
\ee

The field action for a free massless real scalar field on a space-time
modeled as the cylinder is defined by
\be
S[\Phi ]\ =\ \frac{1}{2}I_0 [-\Phi\partial^2_\tau \Phi -
\Phi \partial^2_\varphi \Phi ]\ .
\ee
It can be interpreted as the action describing (one coordinate of the) free
closed bosonic string. We shall have in mind this interpretation in what
follows. We can expand the fields $\Phi$ into Fourier modes
\be
\Phi (x) \ =\ \sum_{k\neq 0} c_k (\tau ) e^{ik\varphi } \ ,\
c^*_k (\tau )\ =\ c_{-k} (\tau )\ ,
\ee
(we eliminated the zero mode which is inessential for us). Inserting (7)
into the action we obtain
\be
S[\Phi ]\ =\ \frac{1}{2}\int_{\bf R} d\tau \ \sum_{k\neq 0} [-c_{-k} (\tau
){\ddot c}_k (\tau )-k^2 c_{-k} (\tau )c_k (\tau )]\ .
\ee
The canonically conjugated momentum to the mode $c_k (\tau )$ is $\pi_k
(\tau )={\dot c}_{-k} (\tau )$. We assume standard equal-time canonical
Poisson brackets between modes and their conjugate momenta
\[
\{ c_k (\tau ),c_{k'} (\tau ) \} \ =\ \{ \pi_k (\tau ),\pi_{k'} (\tau ) 
\} \ =\ 0\ ,
\]
\be
\{ c_k (\tau ), \pi_{k'} (\tau ) \} \ =\ \delta_{kk'}\ .
\ee

The quantization means an operator realization of the corresponding
equal-time canonical commutation relations 
\[
[c_k (\tau ),c_{k'} (\tau )]\ =\ [\pi_k (\tau ),\pi_{k'} (\tau )]\ =\ 0\ ,
\]
\be
[c_k (\tau ),\pi_{k'} (\tau ) ]\ =\ i\delta_{kk'}\ .
\ee
This can be performed by solving corresponding classical Euler-Lagrange
equations of motion
\be
-\ {\ddot c}_k (\tau )\ =\ k^2 \, c_k (\tau )\ .
\ee
They possess positive frequency oscillating solutions
\be
c_k (\tau )\ =\ \frac{i}{k}[a_k e^{-ik\tau} \ -\ b_{-k} e^{ik\tau} ]\ ,
\ k>0\ .
\ee
The solutions for negative $k$ are fixed by the reality condition: $c_{-k}
(\tau )=c^* c_k (\tau )$, i.e. $a_{-k} =a^*_k$, $b^*_k =b_{-k}$.
Explicitly, the formulas for the field $\Phi (\tau ,\varphi )$ and the
conjugated momentum $\Pi (\tau ,\varphi )=\partial_\tau \Phi (\tau ,\varphi
)$ read
\[
\Phi (\tau ,\varphi )\ =\ \sum_{k\neq 0} \frac{i}{k}[a_k e^{-ik\tau +
ik\varphi} \ -\ b_k e^{-ik\tau -ik\varphi} ]\ ,
\]
\be
\partial_\tau \Phi (\tau ,\varphi )\ =\ \sum_{k\neq 0} [a_k e^{-ik\tau
+ik\varphi} \ +\ b_k e^{-ik\tau -ik\varphi} ]\ .
\ee
The terms with expansion coefficients $a_k$ are interpreted as the right-movers 
on a closed bosonic string, whereas those solutions with $b_k$ as the 
left-movers. 
They are independent, and we can treat them separately. The canonical
commutations relations are indeed satisfied if we replace the complex
coefficients $a_k$ and $b_k$, $k\neq 0$, by two independent infinite set of
bosonic oscillators satisfying commutation relations
\be
[a_k ,b_{k'} ]\ =\ 0\ ,\ [a_k ,a_{k'} ]\ =\ [b_k ,b_{k'} ]\ =\
k\delta_{k+k',0}
\ee
(we are using the same notation for the classical expansion parameters and
annihilation and creation operators, this should not lead to any confusion).

\subsection{Noncommutative case}
The noncommutative cylinder we realize by "quantizing" the Poisson
structure on $C$, see \cite{CDP1}-\cite{CDP2}. This is defined by
\be
\{ f,g \} \ =\
\frac{\partial f}{\partial \varphi }\frac{\partial g}{\partial \tau} \ -\
\frac{\partial f}{\partial \tau }\frac{\partial g}{\partial \varphi} \ ,
\ee
for any pair of functions $f$ and $g$ on $C$. Using the Leibniz rule, it can be
generated from the elementary brackets
\be
\{ \tau ,x_\pm \} \ =\ \pm ix_\pm \ ,\ \{ x_+ ,x_- \} \ =\ 0 \ ,
\ee
where, $x_\pm =\rho e^{\pm i\phi}$. Eqs. (17) are just $e(2)$ Lie algebra
defining relations. The function $x_+ x_-$ is central: $\{ x_0 ,x_+ x_- \}
=\{ x_\pm ,x_+ x_- \} =0$, i.e. the restriction $x_+ x_- =\rho^2$ is
consistent with the Poisson bracket structure. The operators
$\partial^2_\varphi$ and $\partial^2_\tau$ entering the action can be
expressed in terms of Poisson brackets:
\be
\partial^2_\varphi f\ =\ \{\tau ,\{\tau ,f\}\} \ ,\ \partial^2_\tau f\ =\
\frac{1}{\rho^2} \{ x_- ,\{\{ x_+ ,f\}\} \ .
\ee

In the noncommutative case we replace the commuting variables $\tau ,
x_\pm$ by the $e(2)$ Lie algebra generators ${\hat \tau},{\hat x}_\pm$
satisfying the relations
\[
[{\hat \tau},{\hat x}_\pm]\ =\ \pm \lambda {\hat x}_\pm \ ,\ 
[{\hat x}_+ ,{\hat x}_- ]\ =\ 0\ ,
\]
\be
{\hat x}_+ {\hat x}_- \ =\ \rho^2 \ .
\ee
The algebra $e(2)$ possesses one series of infinite dimensional unitary
representations (parameterized by one real parameter $\rho >0$). These
representations can be realized in the Hilbert space $L^2 (S^1 ,d\varphi )$
as follows
\be
{\hat \tau} \ =\ -i\lambda \partial_\varphi \ ,\ {\hat x}_\pm \ =\ 
\rho e^{\pm i\varphi} \ .
\ee
The Casimir operator takes the value ${\hat x}_+ {\hat x}_- =\rho^2$. In
what follows we put $\rho =1$.

For any operator $\hat f$ on $C$ possessing the expansion
\be
{\hat f}\ =\ \sum_{k\in{\bf Z}} c_k ({\hat \tau})e^{ik\varphi} \ ,
\ee
we introduce the noncommutative analog of integral (5) by
\be
I_\lambda [{\hat f}]\ =\ \lambda {\rm Tr}[{\hat f}]\ =\ \lambda 
\sum_{n\in {\bf Z}} \ c_0 (n\lambda) \ .
\ee
Here $c_0 (n\lambda)$ are spectral coefficients of the operator $c_0
({\hat \tau})$: $c_0 ({\hat \tau}) e^{in\varphi }= c_0 (n\lambda )$
$e^{in\varphi }$ ($e^{in\varphi }$ is an eigenfunction of ${\hat \tau}$ with 
the eigenvalue $n\lambda$). There is a straightforward generalization of
(22) to an integral over finite discrete time interval $\alpha =a\lambda
\leq\tau\leq b\lambda =\beta$:
\be
I^\beta_{\lambda\alpha} [{\hat f}]\ :=\ \lambda {\rm Tr}^\beta_\alpha [{\hat
f}]\ =\ \lambda \sum^b_a c_0 (n\lambda )\ .
\ee
Integrals of this type appear, e.g., if one calculates the field action for 
a finite time interval.

The operators ${\hat \partial}^2_\varphi$ and ${\hat \partial}^2_\tau$, the
noncommutative analogs of (18), are obtained replacing the Poisson brackets
by the commutators: $\{ .,.\} \to (i/\lambda )[.,.]$. Performing the Fourier
expansion (21) we obtain
\[
{\hat\partial}^2_\varphi {\hat f}\ =\ -\frac{1}{\lambda^2} [{\hat\tau},
[{\hat\tau} ,{\hat f}]]\ =\ -\sum_{k\in {\bf Z}} k^2 c_k ({\hat \tau})
e^{ik\varphi } \ ,
\]
\be
{\hat \partial}^2_\tau {\hat f}\ =\ -\frac{1}{\lambda^2 \rho^2}[{\hat x}_-
,[{\hat x}_+ ,{\hat f}]]\ =\ \sum_{k\in {\bf Z}} \delta^2_\lambda c_k ({\hat
\tau}) e^{ik\varphi } \ .
\ee
Here $\delta^2_\lambda$ is the second order difference operator
\[
\delta^2_\lambda  c_k ({\hat \tau})\ =\ \frac{1}{\lambda^2}[c_k ({\hat 
\tau} +\lambda ) -2c_k ({\hat \tau})+ c_k ({\hat \tau} -\lambda )]\ .
\]

The free hermitian scalar field action we take in the form
\be
S[{\hat \Phi}]\ =\ \frac{1}{2}I_\lambda [-\Phi\partial^2_\tau {\hat\Phi} +
K^2_\lambda (-i\partial_\varphi ){\hat \Phi}]\ ,       \label{pEact}
\ee
where $K_\lambda (-i\partial_\varphi )=\frac{2}{\lambda}\sin
(-i\frac{\lambda}{2} \partial_\varphi )$. This specific form of the operator 
$K_\lambda$ has been chosen for the later convenience: as we shall see in
section~3.2, the Euclidean version of (\ref{pEact}) corresponds precisely to
the action which can be reinterpreted as a theory on the noncommutative
$q$-plane with a most simple and natural Lagrangian.
The operator $K_\lambda
(-i \partial_\varphi )$ is defined by the Fourier expansion (21): $K_\lambda
(-i\partial_\varphi ){\hat f}=\sum c_k ({\hat \tau})K_\lambda (k)e^{ik
\varphi}$;\ $K_\lambda (k)=\frac{2}{\lambda} \sin (\frac{k\lambda}{2})$.

The field ${\hat \Phi}$ is a function in the noncommutative variables ${\hat
\tau}$ and ${\hat x}_\pm$. It possesses, by assumption, the Fourier expansion
\be
{\hat \Phi}\ =\ \sum_{k\neq 0} c_k ({\hat \tau}) e^{ik\varphi } \ ,\quad 
c^*_k ({\hat \tau}) = c_{-k} ({\hat \tau}-k\lambda )\ . 
\ee
The latter relation guarantees the hermiticity of the field:
\[
{\hat \Phi}^* \ =\ \sum_{k\neq 0} e^{-ik\varphi } c^*_k ({\hat \tau})\ =\
\sum_{k\neq 0} c^*_k ({\hat \tau}+k\lambda ) e^{-ik\varphi }
\]
\[
\ =\ \sum_{k\neq 0} c^*_{-k} ({\hat \tau}-k\lambda ) e^{ik\varphi } \ =\
\sum_{k\neq 0} c_k ({\hat \tau}) e^{ik\varphi } \ =\ {\hat \Phi}\ .
\]
Inserting the mode expansion (25) into $S[{\hat\Phi}]$ and using the relation
\be
c_k ({\hat \tau})e^{ik\varphi}c_{k'} ({\hat \tau})e^{ik'\varphi} \ =\
c_k ({\hat \tau}) c_{k'} ({\hat \tau}-k\lambda )e^{i(k+k')\varphi} \ ,
\ee
we obtain
\be
S[{\hat \Phi}]\ =\ \frac{\lambda}{2}\sum_{n,k} [c_{-k} (n\lambda -k\lambda )
\delta^2_\lambda c_k (n\lambda )\ -\ K_\lambda^2 (k)c_{-k} (n\lambda
-k\lambda )c_k (n\lambda )]\ .
\ee
Its extremalization leads to the discrete-time Euler-Lagrange equations
\be
-\delta^2_\lambda c_k (n\lambda )\ =\ K_\lambda^2 (k) c_k (n\lambda )\ .
\ee
We see that the noncommutativity demonstrates itself dominantly as a
discreteness of time: the action (27) does not contain explicitely
noncommutative quantities, it depends on spectral modes at various discrete
time slices. Moreover, in the equations of motion the time derivatives are
replaced by the time differences.

The momentum, conjugated to the field mode $c_k (n\lambda )$, is
\[
\pi_k (n\lambda )\ =\ \frac{1}{2\lambda}[c_k (n\lambda -k\lambda +\lambda
)-c_k (n\lambda - k\lambda -\lambda ) \ .
\]
We postulate the standard equal-time Poisson brackets among modes and
conjugated momenta
\[
\{ c_k (n\lambda ),c_{k'} (n\lambda ) \} \ =\ \{ \pi_k (n\lambda ),
\pi_{k'} (n\lambda ) \} \ =\ 0\ ,
\]
\be
\{ c_k (n\lambda ), \pi_{k'} (n\lambda ) \} \ =\ \delta_{kk'}\ .
\ee
The quantization means an operator realization of the corresponding 
equal-time canonical commutation relations 
\[
[c_k (n\lambda ),c_{k'} (n\lambda )]\ =\ [\pi_k (n\lambda ),
\pi_{k'} (n\lambda )]\ =\ 0\ ,
\]
\be
[c_k (n\lambda ), \pi_{k'} (n\lambda )]\ =\ i\delta_{kk'}\ ,\ .
\ee
This can be performed similarly as in the commutative case, in terms of
suitable sets of annihilation and creation operators. We shall not discuss
this problem here since in the next Section, we analyze in detail the 
analogous problem within the Euclidean version.

\section{Euclidean model on a cylinder}
\subsection{Commutative case}
In order to analyze the Euclidean case (see, e.g., \cite{GSW}, \cite{Pol}), 
we have to continue the time $\tau$
to $-i\tau$, as a result in the action the kinetic term changes sign. The
Euclidean action is usually defined by
\be
S[\Phi ]\ =\ \frac{1}{2}I_0 [-\Phi\partial^2_\tau \Phi
-\Phi\partial^2_\varphi \Phi )]\ .
\ee
The corresponding Euler-Lagrange equations for modes
\be
{\ddot c}_k (\tau )\ =\ k^2 \, c_k (\tau )\ .
\ee
can be solved similarly as in the real-time case. The formula for the field
$\Phi (\tau ,\varphi )$ and the conjugate momentum $\Pi (\tau ,\varphi
)=\partial_\tau \Phi (\tau ,\varphi )$ reads
\[
\Phi (\tau ,\varphi )\ =\ \sum_{k\neq 0} \frac{i}{k}[a_k e^{-k\tau
+ik\varphi} \ +\ b_k e^{-k\tau -ik\varphi} ]\ ,
\]
\be
\partial_\tau \Phi (\tau ,\varphi )\ =\ -i\sum_{k\neq 0} [a_k e^{-k\tau
+ik\varphi} \ +\ b_k e^{-k\tau -ik\varphi} ]\ .
\ee
The equal-time canonical commutation relations among $\Phi (\tau ,\varphi )$
and $\Pi (\tau ,\varphi' )$ are satisfied provided that $a_k$ and $b_k$,
$k\neq 0$, satisfy commutation relations (14).

It is useful to introduce the new complex variables
\be
z= e^{\tau -i\varphi }\ ,\ {\bar z}= e^{\tau +i\varphi }\ .
\ee
In this variables the action reads
\be
S[\Phi ]\ =\ \frac{1}{4\pi}\int dzd{\bar z}\, \partial \Phi (z,{\bar z})
{\bar \partial}\Phi (z,{\bar z})\ .
\ee
The corresponding Euler-Lagrange equations 
\be
\partial {\bar \partial}\Phi (z,{\bar z})\ =\ 0
\ee
have a general solution
\be
\Phi (z,{\bar z})\ =\ i\sum_{k\neq 0} \frac{1}{k}[a_k z^{-k}\ +\ b_k
{\bar z}^{-k} ]\\ =:\ \Phi (z)\ +\ {\bar \Phi}({\bar z})\ .
\ee
Comparing with (33), we see that the fields $\Phi (z)$ and ${\bar \Phi}
({\bar z})$ correspond to the right- and left-movers, respectively, 
of a closed bosonic string.

The energy momentum tensor $T_{ij}$ is traceless: $T_{z{\bar z}} =0$ in
complex notation with $i,j=z,{\bar z}$. This, together with the
energy-momentum conservation,
\[
{\bar\partial}T_{zz} +\partial T_{{\bar z}z} =0\ ,\
\partial T_{{\bar z}{\bar z}} +{\bar\partial}T_{{\bar z}z} =0\ ,
\]
gives $T(z,{\bar z})=T(z)\ +\ {\bar T}({\bar z})$ with
\be
T(z)\ =\ -\frac{1}{2}:\partial \Phi (z)\partial \Phi (z):\ ,\ {\bar
T}({\bar z})\ =\ -\frac{1}{2}:{\bar\partial}{\bar \Phi}({\bar z})
{\bar\partial}{\bar \Phi}({\bar z}):\ .
\ee
Here there appears the normal product $:\dots :$ defined by 
\be
:\Phi (z,{\bar z})\Phi (w,{\bar w}):\ =\ \Phi (z,{\bar z})\Phi (w,{\bar w})
\ +\ \log |z-w|^2 \ .
\ee 
The last term is just the 2-point correlator (Green function) $\langle \Phi
(z,{\bar z})\Phi (w,{\bar w})\rangle$. It can be found, e.g. in the
framework of Euclidean field theory, with the quantum expectation value of
a field functional $F[\Phi ]$ being given as the path integral
\be
\langle F[\Phi ]\rangle \ =\ \int D\Phi \, F[\Phi ]\, e^{-S[\Phi ]}\ .
\ee

Let us now consider the infinitesimal transformation of fields $\delta_\xi
\Phi = \xi \partial \Phi$ with some given function $\xi =\xi (z,{\bar z})$.
This induces the following variation of the action
\[
\delta_\xi S\ =\ \frac{1}{4\pi}\int dzd{\bar z}\, [({\bar\partial} \xi )
(\partial \Phi )^2 +\partial (\xi\, \partial \Phi\, {\bar \partial} \Phi )]
\]
\be
\ =\ -\frac{1}{2\pi}\int dzd{\bar z}\, ({\bar \partial}\xi )T\ ,
\ee
with $T(z)$ given in (38) (here we used the formula $\int dzd{\bar z}\,
\partial (\dots )$ $=0$). The analogous formula for ${\bar T}({\bar z})$ can
be obtained by considering the variations $\delta_{\bar\xi} \Phi ={\bar\xi}
{\bar\partial}\Phi$.

Inserting into $T(z)$ the mode expansion (33), using (38) and replacing the
expansion coefficient by bosonic oscillators, we recover the Sugawara formula:
\be
T(z)\ =\ \sum L_m z^{-m-2} \ ,
\ee
where
\be
L_m \ =\ \frac{1}{2\pi i}\oint z^{n+1} T(z)\ = \frac{1}{2} \sum :a_k a_n :
\delta_{k+n,m} \ .
\ee
We note that here, on r.h.s., there 
appears the standard normal ordering with respect
to the annihilation and creation operators. It can be shown that it is
equivalent to the normal ordering introduced in (39), therefore, we have not
introduced a new specific notation.

For variations $\delta_\xi\Phi =\xi\partial\Phi$ with $\xi =\xi (z)$ it holds
$\delta_\xi S=0$ (see (41)). Thus, they are symmetry transformations of the
model. The variation $\delta_\xi \Phi$ in question corresponds to an
infinitesimal conformal transformation $z\to z+\xi (z)$ of the conformal
plane:
\[
\Phi (z)\ \to \ \Phi (z+\xi (z))\ =\ \Phi (z)+\xi (z)\partial \Phi (z)\ .
\]
The infinitesimal conformal transformations of a plane form a Lie algebra,
and the same is true for the variations: $[\delta_\xi ,\delta_{\xi'} ]\Phi=
\delta_{\xi \partial_{\xi'}- \partial_\xi \xi'} \Phi$. Consequently, the
Virasoro generators $L_m$, $m\in {\bf Z}$, close to a Lie algebra too. Wee
see that the conformal mappings of a complex plane are behind the symmetry
of the action in question.

\subsection{Noncommutative case}
Now we shall investigate along the similar lines the noncommutative
Euclidean version of the model. It is obtained by replacing  $\lambda \to
-i\lambda$ in (27). The corresponding Euclidean action is
\be
S[{\hat\Phi}]\ =\ \frac{1}{2}I_\lambda [-{\hat\Phi}{\hat\partial}^2_\tau
{\hat \Phi} - {\hat\Phi}K^2_\lambda (-i\partial_\varphi ){\hat \Phi}]\ ,
\ee
where now $K_\lambda (-i\partial_\varphi )=\frac{2}{\lambda}\sinh (-i
\frac{\lambda}{2}\partial_\varphi)$. Inserting into (44) the field mode
expansion, we obtain
\be
S[{\hat \Phi}]\ =\ \frac{\lambda}{2}\sum_{n,k} [c_{-k} (n\lambda -k\lambda)
\delta^2_\lambda c_k (n\lambda ) + K^2_\lambda (k) c_{-k} (n\lambda -k\lambda
)c_k (n\lambda )]\ ,
\ee
with $K_\lambda (k)=\frac{2}{\lambda}\sinh (\frac{k\lambda}{2})$. The
corresponding discrete Euler-Lagrange equations
\be
\delta^2_\lambda c_k (n\lambda )\ =\ K^2_\lambda (k)c_k (n\lambda )\ ,
\ k\neq 0\ ,
\ee
have the solution
\be
c_k (n\lambda )\ =\ \frac{i\lambda}{\sinh (k\lambda )} [a_k e^{\lambda
k^2/2} e^{-kn\lambda} - b_{-k} e^{-\lambda k^2/2} e^{kn\lambda} ]\ ,\
k\neq 0\ ,
\ee
where $a_k$ and $b_{-k}$ are independent constants. The conjugate momentum
to the mode $c_k (n\lambda )$ is
\[
\pi_k (n\lambda )\ =\ \frac{1}{2\lambda} [c_{-k} (n\lambda -k\lambda
-\lambda )-c_{-k} (n\lambda -k\lambda +\lambda )]
\]
\be
\ =\ [a_{-k} e^{-\lambda k^2/2} e^{kn\lambda} +b_k e^{\lambda k^2/2}
e^{-kn\lambda} ]\ ,\ k\neq 0\ .
\ee
The Euclidean reality condition, ${\hat\Phi}^\dagger ({\hat\tau},\varphi )
\stackrel{\rm def}{=}{\hat\Phi}^* (-{\hat\tau},\varphi )
={\hat\Phi}({\hat\tau},\varphi )$, requires $c_k (n\lambda)=c^*_{-k} (-n
\lambda +k\lambda)$. Taking this into account, the canonical commutation
relations (30) are satisfied provided that the coefficients $a_k$ and
$b_k$, $k\neq 0$, are replaced by bosonic operators satisfying the commutation
relations
\be
[a_k ,b_{k'} ]\ =\ 0\ ,\ [a_k ,a_{k'} ]\ =\ [b_k ,b_{k'} ]\ =\ 
\frac{\sinh (k\lambda)}{\lambda}\delta_{k+k',0}\ .
\ee

Inserting solution (47) into the field mode expansion, we obtain the field
configuration, minimizing the action, in the form
\be
{\hat \Phi}({\hat z},{\hat{\bar z}})\ =\ \sum_{k\neq 0} \frac{i\lambda}{
\sinh (k\lambda )}[a_k {\hat z}^{-k} - b_k {\hat{\bar z}}^{-k} ]\ =:\
{\hat\Phi}({\hat z})\ +\ {\hat{\bar \Phi}}({\hat{\bar z}})\ .
\ee
Here we introduced the new noncommutative variables   
\be
{\hat z}\ =\ e^{{\hat \tau}-i\varphi} \ =\ e^{-i\lambda \partial_\varphi
-i\varphi} \ ,\ {\hat{\bar z}}=\ e^{{\hat \tau}+i\varphi} \ =\ e^{-i\lambda
\partial_\varphi +i\varphi} \ .
\ee
The operators ${\hat z}$ and ${\hat{\bar z}}$ satisfy the commutation relation
\be
{\hat z}{\hat{\bar z}}\ =\ q^2 {\hat{\bar z}}{\bar z}\ ,\ q=e^\lambda \ .
\ee
which are typical for the $q$-deformed plane ${\bf C}_q$ possessing an
$E_q (2)$ symmetry. Bellow we summarize necessary formulas following the
conventions of \cite{Bon}.

The $q$-deformed Euclidean group $E_q (2)$ is a Hopf algebra generated by
$v$, ${\bar v}$, $n$, ${\bar n}$ with relations
\[
vn\ =\ q^2 nv\ ,\ v{\bar n}\ =\ q^2 {\bar n}v\ ,\ n{\bar n}\ =\ q^2 {\bar
n}n\ ,\]
\[
{\bar n}{\bar v}\ =\ q^2 {\bar v}{\bar n}\ ,\ n{\bar v}\ =\ q^2 {\bar v}n\ =\
v{\bar v}\ =\ {\bar v}v \ =\ 1\ ,
\]
with comultiplication, counit and antipode given by
\[
\Delta v\ =\ v\otimes v\ ,\ \Delta {\bar v}\ =\ {\bar v}\otimes {\bar v}\ ,\
\Delta n\ =\ n\otimes 1 +v \otimes n\ ,\ \Delta {\bar n}\ =\ {\bar n}\otimes
1 + {\bar v}\otimes{\bar n}\ ,
\]
\[
\varepsilon (v)= \varepsilon ({\bar v})=1\ ,\ \varepsilon (n)= \varepsilon
({\bar n})=0\ ,
\]
\[ 
S(v)={\bar v}\ ,\ S({\bar v})=v\ ,\ S(n)=-{\bar v}n\ ,\ S({\bar n})\ =\
-v{\bar n}\ .
\]
For a real $q$ the compatible involution is $v^* ={\bar v}$, $n^* ={\bar n}$.

The dual enveloping algebra ${\cal U}_q (e(2))$ is generated by the elements
$P_\pm$ and $J$ such that
\[
[P_+ ,P_- ]=0\ ,\ [J,P_\pm ]=\pm P_\pm \ ,
\]
with vanishing counit and 
\[
\Delta J=J\otimes 1+1\otimes J\ ,\ \Delta P_\pm =q^{-J} \otimes P_\pm +P_\pm
\otimes q^J \ ,
\]
\[
S(J)=-J\ ,\ S(P_\pm )=-q^{\pm 1} P_\pm \ ,\ J^* =J\ ,\ P^*_\pm =P_\mp \ .
\]

The plane ${\bf C}_q$ is associated with functions analytic in ${\hat z}$,
${\hat{ \bar z}}$. It is invariant with respect to the ${\cal U}_q (e(2))$
coaction $\Lambda$ given by (see \cite{Bon}):
\[
\Lambda (q^{\pm J} ){\hat z}^m {\hat{\bar z}}^n = q^{\pm m\mp n} {\hat z}^m
{\hat{\bar z}}^n \ ,
\]
\[
\Lambda (P_- ){\hat z}^m {\hat{\bar z}}^n \equiv \partial_q {\hat z}^m
{\hat{\bar z}}^n =[m]_q q^{n-2} {\hat z}^{m-1} {\hat{\bar z}}^n \ ,
\] 
\[
\Lambda (P_+ ){\hat z}^m {\hat{\bar z}}^n \equiv {\bar\partial}_q {\hat z}^m
{\hat{\bar z}}^n=-[n]_q q^{m+1} {\hat z}^m {\hat{\bar z}}^{n-1} \ ,
\]
where $[m]_q =(q^m -q^{-m})/(q-q^{-1})$. The corresponding Casimir operator
(the Laplacian on ${\bf C}_q$) is $\Delta_q =\Lambda (P_+ P_-)$. Its action
on ${\bf C}_q$ is
\be
\Delta_q {\hat z}^m {\hat{\bar z}}^n = \Lambda (P_+ P_-){\hat z}^m {\hat
{\bar z}}^n =-q^{m+n-2}[m]_q [n]_q {\hat z}^{m-1} {\hat{\bar z}}^{n-1} \ .
\ee

The operator $\Delta_\lambda =-{\hat \partial }^2_{\bar \tau} +\frac{4}{
\lambda}^2 \sinh^2 (-i\frac{\lambda}{2}\partial_\varphi )$ entering (44)
acts on ${\hat z}^m {\hat {\bar z}}^n = e^{(m+n){\hat\tau}-i(m-n)\varphi -mn
\lambda}$ as follows
\[
\Delta_\lambda {\hat z}^m {\hat{\bar z}}^n=[-{\hat\partial }^2_\tau
+\frac{4}{\lambda}^2 \sinh^2 (-i\frac{\lambda}{2}\partial_\varphi )]
{\hat z}^m {\hat{\bar z}}^n
\]
\be
=-\frac{(q-q^{-1})^2}{ \lambda^2}[m]_q [n]_q {\hat z}^m {\hat{\bar z}}^n
\ ,\ q=e^\lambda \ .
\ee
Comparing (53) and (54) it can be shown straightforwardly that
\be
\Delta_\lambda {\hat z}^m {\hat{\bar z}}^n \ =\ \frac{(q-q^{-1})^2}{
\lambda^2} {\hat r}(\Delta_q {\hat z}^m {\hat{\bar z}}^n ){\hat r}\ ,\
{\hat r}=e^{\hat\tau} \ .
\ee
This is the noncommutative analog of the known link between Laplacian on a
plane and those on a cylinder.
 
To any function $f({\hat z},{\hat{\bar z}})=\sum_{n,m\ge 0} C_{m,n} {\hat
z}^m {\hat{\bar z}}^n$ on ${\bf C}_q$ we assign the Jackson-type integral by
\be
\int_q d{\hat z}d{\hat{\bar z}}\, f({\hat z},{\hat{\bar z}})\ =\ \lambda
{\rm Tr}[{\hat r}^2 f_0 ({\hat r}^2)]\ =\ \lambda\sum_{n\in{\bf Z}} q^{2n}
f_0 (q^{2n}) \ ,
\ee
where ${\rm Tr}$ denotes the trace over the spectrum of ${\hat r}^2$ is
$q^{2n}$, $n\in{\bf Z}$, and
\be
f_0 ({\hat r})\ =\ \sum_{n\ge 0} C_{n,n} q^{n^2} {\hat r}^{2n} \ .
\ee

We can extend the definition (56) by taking a partial trace over the
spectrum: putting $\alpha =q^{2a}$ and $q^b =\beta$, we define
\be
\int^\beta_{q\alpha} d{\hat z}d{\hat{\bar z}}\, f({\hat z},{\hat{\bar z}})\
=\ \lambda \sum^b_{n=a} q^{2n} f_0 (q^{2n}) \ .
\ee
Taking ${\hat f}={\hat r}^2 f({\hat z},{\hat{\bar z}})$ it can be seen easily
that the integrals $I_\lambda [{\hat f}]$ and $\int_q d{\hat z}d{\hat{\bar
z}}\, f({\hat z},{\hat{\bar z}})$ are equal. The factor ${\hat r}^2$
represents a Jacobian of the transformation (51). We see that in the
noncommutative case there are two equivalent approaches, linked by the
transformation (51):

(i) The first one corresponds to the model on a noncommutative sphere
described by the noncommutative variables ${\hat\tau}$ and $\varphi$. The
field action (44) leads to the equations of motion (46).

(ii) The second one is a model on a $q$-plane described by the variables
${\hat z}$, ${\hat{\bar z}}$. The action
\be
S[{\hat\Phi}]\ =\ \int_q d{\hat z}d{\hat{\bar z}}\ {\hat \Phi}({\hat z}q,
{\hat{\bar z}}q^{-1}) (-\Delta_q ){\hat\Phi}({\hat z},{\hat{\bar z}})\ .
\ee
The shifts of arguments in the first ${\hat \Phi}$ guarantee that (59) is
equivalent to (44) (use the link (53) between Laplacians and the relation
${\hat r}{\hat\Phi}({\hat z}q,{\hat{\bar z}}q^{-1})={\hat\Phi}({\hat z},
{\hat{\bar z}}){\hat r}$).

The action (59) depends on general hermitian field configurations
\be
{\hat \Phi}({\hat z},{\hat{\bar z}})\ =\ \sum_{k\neq 0} \frac{i\lambda}{
\sinh (k\lambda )}[a_k (q^k {\hat r}^2 ){\hat z}^{-k} -b_{-k} (q^{-k}{\hat
r}^2 ){\hat{\bar z}}^{-k} ]\ .
\ee
The field is hermitian provided that $a^*_{-k} (q^k {\hat r}^{-2} )=a_k
(q^{-1}{\hat r}^2 )$. The Euler-Lagrange equation for $S[{\hat \Phi}]$ is
just the $q$-harmonicity condition
\be
\Delta_q {\hat\Phi}({\hat z},{\hat{\bar z}})\ =\ \partial_q {\bar\partial}_q
{\hat \Phi}({\hat z},{\hat{\bar z}})\ =\ 0\ .
\ee
Obviously, it is equivalent to the equation of motion (46).

Let us now investigate the variations of the action under infinitesimal
field transformations
\be
\delta_{\hat\xi} {\hat\Phi}\ =\ {\hat\xi}\partial_q {\hat\Phi}\ ,\
{\hat\xi}=\xi ({\hat z},{\hat{\bar z}})\ .
\ee
Using the Leibniz rule
\[
\partial_q [f({\hat z}q,{\hat{\bar z}}q)\, g({\hat z},{\hat{\bar z}})]\ =\
(\partial_q f)({\hat z}q,{\hat{\bar z}}q)\, g({\hat z}q,{\hat{\bar z}}q)\, +\,
f({\hat z}q,{\hat{\bar z}}q)\, (\partial_q g)({\hat z}q,{\hat{\bar z}}q) \ ,
\]
we can rewrite the action in the form
\[
S[{\hat\Phi}]\ =\ q^2 \int_q d{\hat z}d{\hat{\bar z}}\ (\partial_q {\hat
\Phi})({\hat z}q,{\hat{\bar z}}q^{-1})\, ({\bar\partial}_q{\hat\Phi})({\hat
z},{\hat{\bar z}})\ .
\]
It follows straightforwardly,
\[
\delta_{\hat\xi} S[{\hat\Phi}]\ =\ q^2 \int_q d{\hat z}d{\hat{\bar z}}\
[\partial_q ({\hat\xi}\partial_q {\hat\Phi})({\hat z}q,{\hat{\bar z}}q^{-1})
\, ({\bar \partial}_q {\hat \Phi})({\hat z},{\hat{\bar z}})\ ,
\]
\be
\ +\ (\partial_q {\hat\Phi})({\hat z}q,{\hat{\bar z}}q^{-1})\, {\bar
\partial}_q ({\hat\xi}\partial_q {\hat\Phi})({\hat z},{\hat{\bar z}})]\ .
\ee
We shall now restrict ourselves to the vicinity of solutions of the equation
of motion, i.e. we take
\[
{\hat \Phi}({\hat z},{\hat{\bar z}})\ =\ \Phi ({\hat z})\ +\ {\bar\Phi}
({\hat{\bar z}})\ .
\]
The first term in (63) does not contribute since, the integrand can be
written as $\partial_q ({\hat\xi}\partial_q \Phi{\bar \partial}_q {\bar\Phi})$
and $\int_q \, \partial_q (\dots )=0$. Thus,
\be
\delta_{\hat\xi} S[{\hat\Phi}]\ =\ q^2 \int_q d{\hat z}d{\hat{\bar z}}\
(\partial_q \Phi )({\hat z}q,{\hat{\bar z}}q^{-1})\, ({\bar \partial}_q )\xi
({\hat z},{\hat{\bar z}})\, (\partial_q \Phi ({\hat z},{\hat{\bar z}})\ .
\ee 
Any function ${\hat\xi}=\xi ({\hat z},{\hat{\bar z}})$ can be written as a
linear combination of functions ${\hat\xi}_k =\eta ({\hat z}){\hat{\bar z}}^k$.
It holds
\[
\delta_{{\hat\xi}_k} S[{\hat\Phi}]\ =\ \int_q d{\hat z}d{\hat{\bar z}}\
(\partial_q \Phi )({\hat z}q)\, ({\bar \partial}_q {\hat\xi}_k )({\hat z},
{\hat{\bar z}})\, (\partial_q \Phi )({\hat z})
\]
\[
\ =\ q^2 \int_q d{\hat z}d{\hat{\bar z}}\ ({\bar \partial}_q {\hat\xi}_k )
({\hat z},{\hat{\bar z}})\, (\partial_q \Phi )({\hat z}q^{2k+1})\,
(\partial_q \Phi )({\hat z})\ .
\] 
Shifting ${\hat z}\to{\hat z}q^{-k-\frac{1}{2}}$ we obtain for any $k$ the
generator of field transformation in the splitted form
\be
T_k ({\hat z})\ =\ - q^2 :(\partial_q \Phi)({\hat z}q^{k+\frac{1}{2}})\,
(\partial_q \Phi)({\hat z}q^{-k-\frac{1}{2}})\ .
\ee 
This is exactly the formula for the splitted Virasoro generators,
\cite{Sat2}-\cite{LP}. Inserting here the mode expansion (see (50))
\be
(\partial_q \Phi )({\hat z})\ =\ \frac{i\lambda q^{-2}}{(q-q^{-1})}
\sum_{k\neq 0} a_k {\hat z}^{-k-1} \ ,
\ee
we obtain
\be
T_k ({\hat z})\ =\ -\frac{\lambda^2}{(q^2 -1)^2}\sum_{n\in{\bf Z}} L^k_n
{\hat z}^{-n-2} \ .
\ee
Here
\be
L^k_n \ =\ \sum_{l,l'\neq 0} q^{(k-1)(l-l')} :a_l a_{l'} :\delta_{l+l',n} \ ,
\ee
are generators of the (double indexed) deformed Virasoro algebra proposed in
\cite{CP1}:
\be
[L^k_n ,L^{k'}_{n'} ]\ =\ \frac{1}{4}\sum_{\sigma \sigma'} [\frac{n-n'}{2}
+n\sigma' k' +n'\sigma k]_- L^{\sigma k'-\sigma' k}_{n+n'} \ +\ C^{kk'}_n
\delta_{n+n',0} \ ,
\ee
where
\be
C^{kk'}_n \ =\ \frac{1}{2}\sum_{m=1}^n [(n-2m)k]_+ [(n-2m)k']_+ [m]_-
[n-m]_- \ .
\ee
Here we introduced the notation $[x]_- =(q^x -q^{-x})/(q-q^{-1} )$ and
$[x]_+ =(q^x +q^{-x})/2$.

Some comments are in order:

(i) In the commutative case the formula (42) for the transformation
generators $T(z)$ follows for a general $\xi (z,{\bar z})$ and a general
field configuration. In addition, in the commutative version,  the
model has the conformal symmetry.

(ii) In the noncommutative case the situation is different: we have
different expressions for $T_k ({\hat z})$ for integers $k$ (depending on 
the point splitting of the arguments in (65)) which generate transformations 
of $q$-harmonic functions among each other. They form the deformed Virasoro
algebra introduced in \cite{CP1} (a particular realization of the
Zamolodchikov-Faddeev algebra \cite{LP}). Notice that the appearance of the 
additional index $k$ in the deformed Virasoro algebra is directly 
related to the noncommutativity of the underlying two-dimensional space-time.

(iii) There are various reasons for which  these symmetries of the model can
not have  as a background some "conformal symmetry" of a $q$-plane: 1)
transformations of the type $z\to z+\xi (z)$ do not spoil the ${\bf C}_q$
structure only for a very limited set of $\xi (z)$, and 2) the simple
$q$-Taylor expansion formula $\Phi (z+\xi (z))=\Phi (z)+\xi (z)\partial_q \Phi
(z)$ $+\dots$, is not valid even for an infinitesimal $\xi (z)$. The meaning of
symmetry transformations, generated by the deformed Virasoro operators,
requires further investigations. This problem is currently under study.

\section{Concluding remarks}
Recently have been found deep relations between string theory and
noncommutative geometry in the space-time \cite{SW}, \cite{Scho}. We have
analyzed an alternative possibility introducing the noncommutative
geometry on a string world-sheet. To achieve this aim we have investigated
a free bosonic string on a noncommutative cylinder. Our results can be
summarized as follows:

- The field theory on a noncommutative cylinder leads in a natural way to
the discrete time evolution \cite{CDP1}-\cite{CDP2}. We started with a
suitable model for a free scalar field (the one component of bosonic
string) on a noncommutative cylinder with a suitable particular symmetry of
the field action.

- The model in the Euclidean version can be equivalently formulated as a
model on a $q$-deformed complex plane ${\bf C}_q$. Its symmetry is
described by the deformed Virasoro algebra suggested earlier \cite{CP1},
appearing in the context of Zamolodchikov-Faddeev algebras \cite{LP}.

The field theoretical origin  of the deformed Virosoro algebra can serve for a
better (physical) motivation and  understanding of its role in  all related
constructions ($q$-strings, $q$-vertex operators and  Zamolodchikov-Faddeev
algebras). We see that the suggested formal deformation of the Virasoro algebra
appears not only within the developed mathematical structure of
Zamolodchikov-Faddeev algebras but also  that there exists a physical theory,
namely,  the free bosonic string on a noncommutative cylinder, in which
framework the deformed Virasoro algebra emerges as the symmetry of the theory.
This fact allows one to justify the appearance and to  clarify the meaning of
the second index which labels (in addition to the usual one) the deformed
Virasoro generators. As we have shown, this is related to the noncommutativity
of the underlying two-dimensional space (world-sheet).    

In this context it would be of great interest to extend our model to the
supersymmetric case. We have strong indications that this can be achieved along
the same lines as in the bosonic case:

  (i) One can start from the fermionic realization of the deformed Virasoro
algebra (69) proposed in \cite{BC}. Introducing the Dirac operator on a
noncommutative cylinder this can be interpreted in terms of a fermionic
spinor field on a noncommutative cylinder.

  (ii) Such spinor model can be reformulated as a theory on a
noncommutative supercylinder. Consequently, the bosonic and fermionic
realizations can be joined to a superfield theory on a noncommutative
supercylinder. In the Euclidean version the resulting theory is formulated
on a $q$-deformed superplane (with symmetries described by the quantum
supergroup $s$-$E_q (2))$.

   Both indicated steps require careful constructions of all noncommutative
analogs of objects in question (the noncommutative (super)cylinder,
Dirac operator, operator orderings, etc). Investigations in this direction
are under current study.

\vspace{5mm}

{\bf Acknowledgments} 

The financial support of the Academy of Finland under the Projects No. 163394 
is greatly acknowledged. 
A.D.'s work was partially supported also by RFBR-00-02-17679 grant
and P.P.'s work by VEGA project 1/4305/97.


\begin{thebibliography}{99}
\bibitem{GSW} M. B. Green, J. H. Schwarz and E. Witten, {\it Superstring
Theory}, Cambridge University Press 1987.
\bibitem{Pol} J. Polchinski, {\it An Introduction to the Bosonic String},
Cambridge University Press 1994.
\bibitem{Con} A. Connes, {\it Noncommutative Geometry}, Academic Press 1994.
\bibitem{CR} A. Connes and M. Reiffel, {\it Yang-Mills for Noncommutative
Two-Tori} in Operator Algebras and Mathematical Physics (Iowa City, Iowa
1985) pp 237, Contemp. Math. Oper. Alg. Math. Phys. 62, AMS 1987.
\bibitem{CDS} A. Connes, M. R. Douglas and A. Schwarz, {\it Noncommutative
Geometry and Matrix Theory: Compactification on Tori}, JHEP 9802:003 (1998),
hep-th/9711152.
\bibitem{SW} N. Seiberg and E. Witten, {\it String Theory and Noncommutative
Geometry}, hep-th/9908142; JHEP 9909: 032 (1999)
\bibitem{Scho} V. Schomerus, {\it D-Branes and Deformation Quantization},
JHEP 9906: 030 (1999), hep-th/9903205.
\bibitem{Kon} M. Kontsevich, {\it Deformation Quantization of Poisson
Manifolds, I}, hep-th/9709040.
\bibitem{CF} A. S. Caetano and G. Felder, {\it A Path Integral Approach to
the Kontsevich Quantization Formula}, math.QA/9902090.
\bibitem{CP1} M. Chaichian and P. Pre\v{s}najder, Phys. Lett. B 277 (1992) 109.
\bibitem{BC} A. Belov and K. D. Chalkitian, Mod. Phys. Lett. A 13 (1993).
\bibitem{Sat1} H. Sato, Nucl. Phys. B 393 (1993) 442.
\bibitem{OS} C. Oh and K. Singh, preprint NUS/HEP/94203 (1994), 
hep-th/9408001.
\bibitem{Sat2} H.-T. Sato, Nucl. Phys. B 471 (1996) 553.    
\bibitem{CP2} M. Chaichian and P. Pre\v{s}najder, Nucl. Phys. B 482 (1996) 466.
\bibitem{LP} S. Lukyanov and Ya. Pugai, {\it Bosonization of ZF algebras:
Direction toward deformed Virasoro Algebra}, hep-th/9412128; JETP 82 (1996)
1021.
\bibitem{Od} S. Odake, {\it Beyond CFT: Deformed Virasoro and elliptic
algebras}, Lect. at 9th CRM Summer School (1999);  hep-th/9910226.
\bibitem{CDP1} M. Chaichian, A. Demichev and P. Pre\v{s}najder, {\it
Quantum field theory on noncommutative space-times and the persistence of
ultraviolet divergences}, hep-th/9812180;  Nucl. Phys. B567 (2000) 360.
\bibitem{GP} H. Grosse and P. Pre\v{s}najder, Acta Phys. Slov. 49 (1999) 185.
\bibitem{CDP2} M. Chaichian, A. Demichev and P. Pre\v{s}najder, {\it
Quantum field theory on the noncommutative plane with $E_q (2)$ symmetry},
hep-th/9904132;  J. Math. Phys. 41 (2000) 1647
\bibitem{Bon} F. Bonechi, N. Ciccoli, R. Giachetti, E. Sorace and M. Tarlini,
Commun. Math. Phys. 175 (1996) 161.
\end{thebibliography}
\end{document}